\documentclass[showpacs,amsmath,
preprintnumbers,
twocolumn,
superscriptaddress,
aps,prb]{revtex4}
\usepackage{graphicx}
\usepackage{dcolumn}

\begin{document}

\title{Optoelectronic control of spin dynamics at near-THz frequencies in magnetically doped quantum wells}
\author{R. C. Myers}
\affiliation{Center for Spintronics and Quantum Computation, 
University of California, Santa Barbara, CA 93106.}
\author{K. C. Ku}
\affiliation{Department of Physics and Materials Research Institute, 
The Pennsylvania State University, University Park, Pennsylvania 16802.}
\author{X. Li}
\affiliation{Department of Physics and Materials Research Institute, 
The Pennsylvania State University, University Park, Pennsylvania 16802.}
\author{N. Samarth}
\affiliation{Department of Physics and Materials Research Institute, 
The Pennsylvania State University, University Park, Pennsylvania 16802.}
\author{D. D. Awschalom}
\affiliation{Center for Spintronics and Quantum Computation, 
University of California, Santa Barbara, CA 93106.}
\date{\today}
\begin{abstract}
We use time-resolved Kerr rotation to demonstrate the optical and electronic tuning of 
both the electronic and local moment (Mn$^{2+})$ spin dynamics in electrically gated 
parabolic quantum wells derived from II-VI diluted magnetic semiconductors. By changing 
either the electrical bias or the laser energy, the electron spin precession frequency is varied 
from 0.1 to 0.8 THz at a magnetic field of 3 T and at a temperature of 5 K. The corresponding 
range of the electrically-tuned effective electron g-factor is an order of magnitude larger 
compared with similar nonmagnetic III-V parabolic quantum wells. Additionally, we demonstrate
that such structures allow electrical modulation of local moment dynamics in the solid state,
which is manifested as changes in the amplitude and lifetime of the Mn$^{2+}$ spin precession signal
under electrical bias. The large variation of electron and Mn-ion spin dynamics is explained 
by changes in magnitude of the $sp-d$ exchange overlap.
\end{abstract}
\pacs{75.50.Pp, 78.47.+p, 85.35.Be, 71.70.Gm}
\maketitle

The drive towards quantum information processing using semiconductors has 
motivated several recent experiments that use high sensitivity time-resolved 
optical techniques to demonstrate the electrical control of coherent 
electron spin dynamics in non-magnetic III-V semiconductor 
heterostructures.\cite{Awschalom:2002} Proof-of-concept experiments have 
exploited variations in the spin-orbit interaction due to quantum 
confinement in a heterostructure,\cite{Poggio:2004} compositional 
gradients,\cite{Salis:2003} and epitaxial strain. \cite{Kato:2004} The 
energy of these effects in conventional semiconductors, however, is on the 
$\mu $eV scale and is manifested as variations in the spin precession 
frequency in the GHz range. In contrast, THz electron spin precession 
frequencies are easily achieved in magnetic semiconductors because the 
Hamiltonian includes the $s-d$ exchange interaction which leads to giant spin 
splittings in the conduction band states on the 10 - 100 meV energy 
scale.\cite{Dietl:1999} Demonstration of electrically-tunable spin 
precession in the GHz-THz regime would provide an important advance towards 
the addressing of quantum states at higher frequencies than currently 
possible in conventional semiconductors. 

Previous work has demonstrated that magnetic doping of II-VI quantum wells 
(QWs) can be used to engineer the exchange overlap between electron/hole 
wave functions and local moments (Mn$^{2+}$ ions).\cite{Crooker:1999} 
This leads to large enhancements of the electron spin precession frequencies 
and the observation of Mn$^{2+}$ spin precession.\cite{Crooker:1996} 
Recently, photoluminescence (PL) measurements in conventional III-V QWs with 
a proximal ferromagnetic barrier have also revealed that a bias voltage may 
be used to tune the spin coupling between carriers in the QW and the 
ferromagnetic layer; however spin dynamics could not be measured in these 
structures.\cite{Myers:2004} Here, we demonstrate the optoelectronic 
tuning of electron spin precession frequency from the GHz to near-THz regime 
in electrically biased, magnetically-doped II-VI parabolic quantum wells 
(PQWs). Electrical bias also modifies the local moment (Mn$^{2+})$ spin 
dynamics indicating a change in the magnetic-ion and hole exchange overlap. 
Large variations in the electron spin dynamics occur as a function of the 
excitation laser energy due to changes in the occupation of states in the 
PQWs. Tuning of the $sp-d$ exchange overlap between the Mn$^{2+}$ and the electron and hole wave functions is therefore demonstrated using both
electrical bias and laser excitation energy.

The samples are grown by molecular beam epitaxy (MBE) and consist of 100 nm 
ZnSe /50 or 100 nm PQW/500 nm ZnSe/ (001) n+GaAs wafer. The optically active 
region consists of a ZnSe-Zn$_{0.85}$Cd$_{0.15}$Se PQW in which the 
concentration of Cd is graded using digital shuttering with a 2.5 nm period 
for the 50 nm sample (Sample 1) and a 5 nm period for the 100 nm sample in 
order to achieve a parabolic band edge profile,\cite{Miller:1984} as shown 
schematically in Fig. 1a. Mn doping is performed using digital shuttering 
such that four 1/8 monolayers of MnSe are deposited within a 5 nm region at 
the center of each well. Control samples of both 50 nm width (Sample 2) and 
100 nm width containing no Mn doping are also grown. A vertical electrical 
bias ($V_{b}$) is applied between a transparent front gate of evaporated 
Ti/Au and the n+GaAs substrate.

\begin{figure}[b]\includegraphics{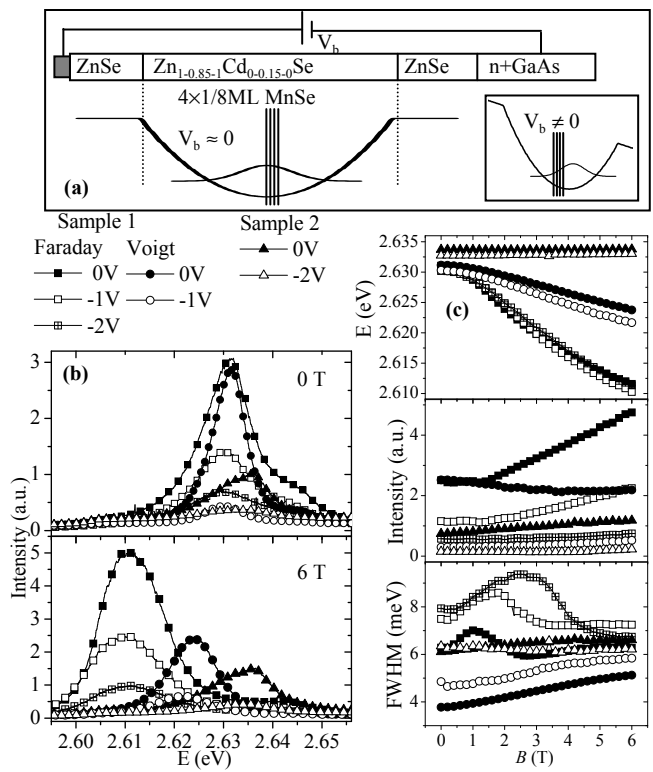}\caption{\label{fig1}
(a) Schematic of sample structure (not to scale), electrical wiring, 
conduction band edge and ground state electron wave function. The inset 
illustrates wave function spatial translation under vertical bias. (b) 
Photoluminescence of 50 nm Mn-doped (Sample 1) and 50 nm control (sample 2) 
at 5 K and $B=0$ T (top) and 6 T (bottom) under various $V_{b}$. (c) PL 
emission energy (E), intensity, and FWHM extracted from Gaussian fits to PL 
as a function of E.
}\end{figure}

PL is measured normal to the sample surface as a function of magnetic field 
($B$) in both Voigt (optical axis $\bot\ B$) and Faraday (optical axis // $B$) 
geometries under fixed vertical bias (Fig. 1b). The excitation consists of 
linearly polarized light from a pulsed laser at an energy of 2.88 eV, power 
density of 56 W/cm$^{2}$, and with a $\sim50\ \mu$m average spot 
diameter. Gaussian fits to the quantum well PL peaks are used to extract 
intensity, emission energy, and linewidth as a function of magnetic field 
for a variety of vertical biases (Fig. 1c). The Mn-doped 50 nm PQW (sample 
1) shows a large Zeeman shift of $\sim20$ meV for magnetic fields between 0 
T and 6 T in the Faraday geometry and $\sim10$ meV in the Voigt geometry, 
whereas the corresponding control (sample 2) shows $<0.3$ meV shift. The 
larger shift in Faraday versus Voigt geometry for the magnetic sample is 
attributed to the lifting of the valence band degeneracy due to quantum 
confinement. In the Faraday case, heavy hole ($j=3/2$) spins are pinned out of 
plane leading to larger spin splitting than in the Voigt 
geometry.\cite{Martin:1990} We have observed PL energy shifts of up to 2 
meV due to $V_{b}$ in both the magnetic and non-magnetic PQW (compare closed 
symbols to open symbols in Fig. 1c). However, precision measurement of these 
shifts is hampered by the rapid quenching of the PL intensity and the 
linewidth broadening observed under vertical bias.

Spin dynamics are measured by time-resolved Kerr rotation (KR) in the Voigt 
geometry with the optical path parallel to the growth 
axis.\cite{Crooker:1996} A Ti:Sapphire laser with a 76 MHz repetition rate 
and $\sim150$ fs pulse width is split into a 1 mW pump beam and a 0.2 mW 
probe beam. The helicity of the pump beam polarization is modulated at 50 
kHz by a photo-elastic modulator, while the intensity of the linearly 
polarized probe beam is modulated by optical chopper at 1 kHz for lock-in 
detection. The pump and probe beams are focused to an overlapping $\sim50 \mu$m diameter spot on the sample, which sits in a magneto-optical 
cryostat with a split-coil superconducting magnet. 

We measure the rotation of the axis of polarization ($\theta _{K}$) of the 
reflected probe beam as a function of time ($\Delta t$) using an optical 
delay line in the path of the probe beam. In this geometry, angular momentum 
is injected parallel to the sample growth axis and perpendicular to the 
magnetic field at $\Delta t$ = 0, resulting in a spin precession about the 
applied field. In magnetically doped samples, both a short lived electron 
spin precession and a long lived Mn$^{2+}$ spin precession are observed. 
The latter precession originates from an effective tipping field due to the 
$p-d$ interaction between the short-lived photo-injected spin-polarized holes and 
the paramagnetically aligned Mn$^{2+}$ spin.\cite{Crooker:1996} The 
data can be fit to two exponentially decaying cosines with the Kerr rotation 
expressed as, $\theta _{K}(\Delta t)=A_{e} e^{( - \Delta 
t/T_{2e}^{\ast })}cos(2\pi \nu _{e}\Delta t + {\phi }_{e}) + 
A_{Mn} e^{( - \Delta t/T_{2Mn}^{\ast })}cos(2\pi \nu 
_{Mn}\Delta t + {\phi }_{Mn})$, where $A_{e}$ ($A_{Mn}$) is the amplitude of 
electron (Mn$^{2+}$) spin precession, $T_{2e}^{\ast }$ 
($T_{2Mn}^{\ast }$) is the transverse electron (Mn$^{2 + }$) spin lifetime, 
$\nu _{e}$ ($\nu _{Mn}$) is the electron (Mn$^{2 + }$) spin 
precession frequency, and $\phi_{e}$ ($\phi_{Mn}$) is a phase offset. The precession 
frequencies are expressed in terms of effective g-factors, $\nu _{e} 
= g_{e}\mu _{B}B/h$, and $\nu _{Mn}=g_{Mn}\mu _{B}B/h$, where 
$g_{e}$ and $g_{Mn}$ are the effective g-factors for electron and Mn$^{2+}$ spin respectively, $\mu _{B}$ is the Bohr magneton, and $h$ is Planck's 
constant.

Figure 2a plots Kerr rotation as a function of both vertical bias and time 
delay for Sample 1 at 5 K and $B$ = 3 T with a pump-probe energy ($E_{p}$) of 
2.638 eV. Two distinct spin precession components are observable and the 
parameters extracted from fits to the data (Fig. 2b) are presented in Fig. 
2c-e. The short-lived component is attributed to electron spin, and reveals 
large tuning of the precession frequency ($g_{e}$ = 1.6 to 8.4) with vertical 
bias, while the long-lived component is attributed to Mn$^{2+}$ spin 
precession whose frequency remains constant ($g_{Mn}=2.0240\pm0.0006$ 
fit precision, not systematic error).

\begin{figure}\includegraphics{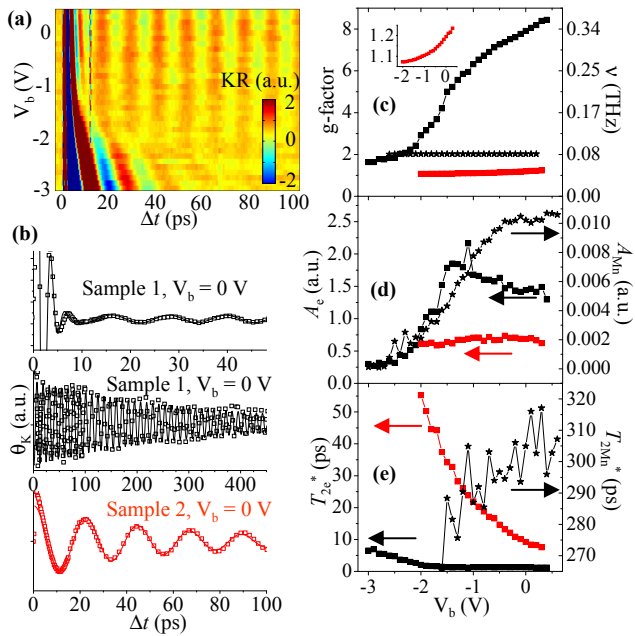}\caption{\label{fig2}
(color) (a) KR of sample 1 as a function of $V_{b}$ and $\Delta t$ at 5 K and 
$B=3$ T. (b) KR as a function of $\Delta t$ at fixed $V_{b}$. Raw data 
(squares) and fits (lines) are plotted in three panels: (top) sample 1 at 
short $\Delta t$ fit to two decaying cosines for both electron and Mn-ion 
spin precession, (middle) sample 1 at long $\Delta t$ fit to a single 
decaying cosine for Mn spin precession only, and (bottom) sample 2 fit to a 
single decaying cosine for electron spin precession only. (c) Spin 
precession frequency and effective g-factor (inset: sample 2 data enlarged 
to show non-magnetic g-factor tuning due to Cd concentration gradient), (d) 
amplitude, and (e) lifetime are plotted as a function of $V_{b}$. Sample 1 
data are black points with squares (stars) for electron (Mn) parameters and 
sample 2 data are red squares.
}\end{figure}

Several observations indicate that a spatial translation of the electron 
wave function due to vertical bias occurs in both the magnetic and 
non-magnetic samples. The non-magnetic sample (sample 2) reveals the 
expected g-factor tuning due to the Cd concentration gradient (inset of Fig. 
2c) and an increase in $T_{2e}^{\ast }$ under negative bias. Both these 
effects have been observed in non-magnetic GaAs-AlGaAs III-V PQW, where the 
g-factor tuning was attributed to the spatial translation of the electron 
wave function into regions of increased Al concentration and the increase in 
lifetime was attributed to the electron and hole separation (splitting of 
the exciton) with bias.\cite{Salis:2003} For the magnetic sample at zero 
bias, where we would expect maximum overlap with the Mn layers, we observe 
the maximum electron g-factor and minimum $T_{2e}^{\ast }$, while negative 
bias leads to a decrease in $g_{e}$ and an increase in $T_{2e}^{\ast }$. 
These changes reflect the fact that $s-d$ overlap both increases the electron 
spin splitting and enhances electron spin scattering, noting that 
$T_{2e}^{\ast }$ is larger in the non-magnetic sample over the bias range 
measured. Qualitatively identical vertical biasing effects are observed in 
another Mn-doped 50 nm PQW with 1 nm period thickness, and in the 100 nm 
Mn-doped PQW, where the effective electron g-factor tunes from 2 to 14 (data 
not shown), thus demonstrating the reproducibility of $s-d$ enhanced g-factor 
tuning in II-VI PQW.

The vertical bias also leads to a spatial translation of the hole wave 
function, and is manifested by changes in the Mn$^{2+}$ spin dynamics. At 
zero bias, the Mn$^{2+}$ spin precession amplitude ($A_{Mn})$ and the spin 
lifetime ($T_{2Mn}^{\ast }$) are at a maximum, while negative bias 
leads to a decrease in both quantities. As mentioned earlier, the Mn$^{2+}$ spin precession is triggered by an effective magnetic tipping pulse 
applied to the paramagnetically aligned Mn$^{2+}$ spin by the short-lived 
spin-polarized holes that are injected at $\Delta t$ = 0. At zero bias, the 
Mn$^{2+}$ spin precession amplitude is large since the hole exchange 
($p-d$) overlap and, therefore, the effective tipping field is large; under 
vertical bias the $p-d$ overlap is decreased such that the tipping field and, 
therefore, $A_{Mn}$ both decrease. We speculate that the decrease in Mn$^{2+}$ spin lifetime with vertical bias could be due to an increase in the 
recombination time of photo-excited carriers as a result of splitting the 
exciton. If the recombination time increases then the temporal cross-section 
for carrier-Mn spin scattering increases.

We also examine the dependence of the electron spin coherence to the laser 
pump-probe energy ($E_{p}$) and find that the electron spin precession 
frequency is maximized at the lowest $E_{p}$ ($g_{e}\sim18$ at 2.611 eV), 
while it saturates for higher $E_{p}$ ($g_{e}\sim6$ at 2.627-2.650 eV), 
as shown in Fig. 3. Given the magnitude of the electron g-factor shifts due 
to $E_{p}$ in comparison to the g-factor tuning due to vertical bias (Fig. 
2c), the $E_{p}$ effect is interpreted as a result of the dependence of the 
electron and Mn overlap (at the center of the wells) on the sublevel energy 
within the PQW. Qualitatively, the s-like ground state wave function (probed 
under the lowest energies) achieves a maximum overlap integral with the 
Mn$^{2+}$ doped region, whereas excited state wave functions have more 
widely distributed probability densities and thus overlap less with Mn$^{2+}$ in the center of the wells. This is consistent with the decrease in 
electron effective g-factor (Fig. 3c) and increase in electron spin lifetime 
(Fig. 3e) observed under higher pump-probe energy. The electron spin 
precession amplitude has sharp features when $E_{p}$ is resonant with the 
heavy hole exciton lines,\cite{Crooker:1995} which are labeled as $n=1$ 
(red line) and $n=2$ (blue line) in Fig. 3d. These features match the 
qualitative interpretation described above such that an increasing $E_{p}$ 
leads to photo-excited electron occupation of both $n=1$ and $n=2$ states 
resulting in a decreasing $g_{e}$. Once $E_{p}$ is at or above the $n=2$ 
energy, $g_{e}$ saturates because occupation and measurement of $n=1$ states 
is unlikely. A quantitative understanding of this effect would require 
two-color pump-probe measurements.

\begin{figure}\includegraphics{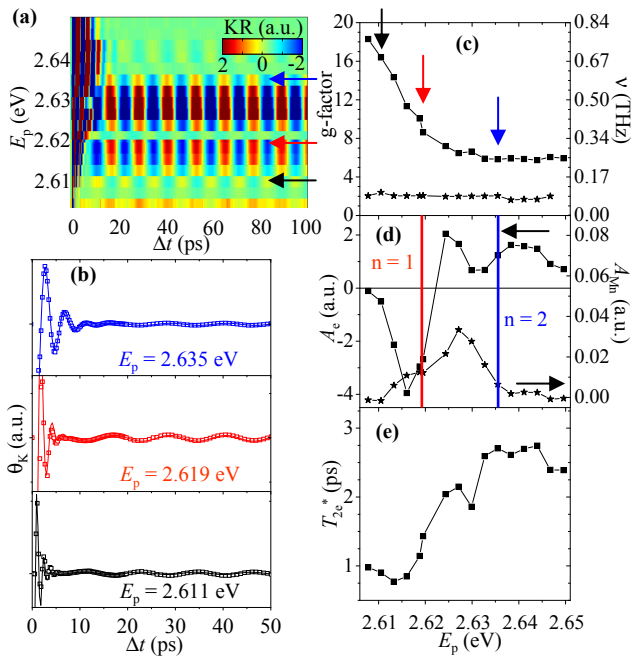}\caption{\label{fig3}
(color) (a) KR at 5 K and $B=3$ T of sample 1 as a function of $E_{p}$. (b) 
Line cuts at constant $E_{p}$ of data from (a), marked by colored arrows, are 
plotted as a function of $\Delta t$. Both raw data (squares) and fits (lines) 
are presented. (c) Spin precession frequency and effective g-factor, (d) 
amplitude, and (e) lifetime are plotted as a function of $E_{p}$. In part (d) 
vertical colored lines label features corresponding to heavy hole exciton 
resonance lines for the ground state ($n=1$, red line) and first excited 
state ($n=2$, blue line).
}\end{figure}

Due to the large pump-probe energy dependence of the electron spin coherence 
as presented in Fig. 3, the vertical biasing effects observed for both 
electron and Mn$^{2+}$ spin coherence in magnetic PQW (Fig. 2) could be 
interpreted, not as a result of spatial wave function translation as 
previously discussed, but as due to the quantum confined stark 
effect\cite{Miller:1985} (QCSE) redshift of the absorption edge such that, 
for a given $E_{p}$, the application of a vertical bias would lead to higher 
energy sublevels being probed and therefore a decrease in the effective 
electron g-factor (Fig. 3). However, a correspondingly large QCSE shift is 
not observed in the gated PL data (Fig. 1) in which under applied bias we 
observe a maximum shift of 2 meV in the PL emission energy, while the 
application of the same bias results in an almost ten-fold decrease in the 
electron precession frequency. We also note that the largest change in 
$g_{e}$ is observed over a $\sim20$ meV pump-probe energy range, thus QCSE 
shifting ($\leq2$ meV) is unlikely to be the cause of the change in $g_{e}$ under 
vertical bias.

Our data indicate that both vertical biasing and choice of pump-probe energy 
can be used to alter electron and Mn$^{2+}$ spin dynamics in a magnetic 
II-VI PQW. As expected, the large exchange coupling between electron and 
Mn$^{2+}$ spin in these structures increases the electron g-factor tuning 
by at least an order of magnitude compared with non-magnetic II-VI or III-V 
PQW structures. A change in magnitude of the $sp-d$ exchange overlap has been 
observed to occur due to the spatial translation of the carrier wave 
functions under vertical bias. The change in electron spin dynamics due to 
laser energy is also interpreted in the context of exchange overlap, such 
that higher laser energy probes spatially broader carrier wave functions. In 
the current symmetric Mn doping scheme, the vertical bias pulls both 
electron and hole wave functions away from the Mn ions; future structures 
with asymmetric Mn doping could be envisaged in which the effects of $s-d$ and 
$p-d$ exchange interaction on the spin dynamics could be separately examined. 

\begin{acknowledgments}
We thank M. Poggio, Y. K. Kato, J. Berezovsky, A. W. Holleitner, and R. J. 
Epstein for a critical reading of the manuscript. Work supported by grants 
NSF-DMR-0305223 and -0305238, DARPA/ONR N0014-99-1-1096 and -1093, AFOSR 
F49620-02-10036, and ARO DAAD19-01-1-0541.
\end{acknowledgments}

\end{document}